\documentclass[twocolumn,pre,superscriptaddress]{revtex4}

\usepackage{dcolumn}
\usepackage{amsmath}

\usepackage{graphicx}
\usepackage{subfigure}

\newcommand{\half}{\mbox{$\frac12$}}
\newcommand{\set}[1]{\lbrace#1\rbrace}
\newcommand{\av}[1]{\langle#1\rangle}
\newcommand{\etal}{{\it{}et~al.}}
\newcommand{\defn}{\textit}

\begin{document}

\title{Robustness of community structure in networks}
\author{Brian Karrer}
\affiliation{Department of Physics, University of Michigan, Ann Arbor, MI
48109}
\author{Elizaveta Levina}
\affiliation{Department of Statistics, University of Michigan, Ann Arbor,
  MI 48109}
\author{M. E. J. Newman}
\affiliation{Department of Physics, University of Michigan, Ann Arbor, MI
48109}
\affiliation{Center for the Study of Complex Systems, University of
Michigan, Ann Arbor, MI 48109}

\begin{abstract}
  The discovery of community structure is a common challenge in the
  analysis of network data.  Many methods have been proposed for finding
  community structure, but few have been proposed for determining whether
  the structure found is statistically significant or whether, conversely,
  it could have arisen purely as a result of chance.  In this paper we show
  that the significance of community structure can be effectively
  quantified by measuring its robustness to small perturbations in network
  structure.  We propose a suitable method for perturbing networks and a
  measure of the resulting change in community structure and use them to
  assess the significance of community structure in a variety of networks,
  both real and computer generated.
\end{abstract}

\maketitle

\section{Introduction}
\label{sec:intro}
Many networks of scientific interest decompose naturally into communities
or modules, densely connected subsets of nodes with only sparser
connections between them.  In many cases communities have been found to
correspond to behavioral or functional units within networks, such as
functional modules in biochemical networks or social groups within social
networks.  This finding suggests that in networked systems whose function
is less well understood we may be able to gain insight by discovering and
examining their communities (if any), and methods for community discovery
have, as a result, attracted a substantial amount of attention in the
recent literature in many disciplines~\cite{Newman04b,Danon2005}.

Communities are of interest for other reasons as well.  Their presence can,
for example, dramatically alter the behavior of dynamical processes on
networks~\cite{ADP06} (and indeed the observation of dynamical processes
has been proposed as one possible method of community
detection~\cite{Boccaletti2007}).  Communities can also be used as a basis
for the reduction or coarse-graining of networks for visualization or other
purposes~\cite{NG04,Arenas2007}.  And communities frequently display
different statistics from the network as a whole, indicating that global
network statistics such as degree moments or correlation functions may
potentially fail to register important heterogeneities~\cite{Newman2006}.

A large number of methods for finding communities have been proposed in
recent years, including divisive methods based on betweenness and similar
measures~\cite{GN02,Radicchi04}, methods based on searching for small
cliques~\cite{Derenyi2005,Palla2007}, information-theoretic
techniques~\cite{Rosvall2007}, statistical inference through belief
propagation~\cite{Hastings2006} or maximum likelihood~\cite{CNM06}, and
many others.

Perhaps the most widely used technique, however, is the maximization of the
benefit function known as
modularity~\cite{NG04,Newman04a,CNM04,DA05,GA05,MAD05,RB06,Newman06b},
which is (to within a multiplicative constant) the difference between the
number of edges within communities and the expected number of such edges
under an appropriate null model.  Various null models have been used but
the commonest by far is the standard configuration
model~\cite{Luczak92,MR95}, which preserves the degree sequence of the
original network but otherwise randomizes edge positions.  The modularity
is then maximized over possible divisions of the network, the optimal
division being taken to be the correct partition of the network into
communities.

Unfortunately, exhaustive maximization of the modularity is known to be an
NP-complete task~\cite{Brandes07} and hence is essentially intractable for
all but the smallest of networks.  In practical implementations of the
modularity method, therefore, approximate heuristics are usually employed,
such as greedy algorithms~\cite{Newman04a,CNM04}, extremal
optimization~\cite{DA05}, simulated annealing~\cite{GA05,MAD05,RB06}, or
spectral methods~\cite{Newman06b}.  These methods vary in their
effectiveness and speed, the faster algorithms tending to give poorer
results while the slower ones can only be applied to smaller networks if
running time is to be kept to reasonable levels.  In this paper we employ
the spectral optimization method introduced in~\cite{Newman06b}, which
displays a reasonable balance between accuracy and speed, but the
calculations we describe are not tied to this method, or even to modularity
maximization in general, and could be applied to any community detection
scheme with only minor modifications.

Despite the large volume of work on community detection and its
applications, one important question remains largely unaddressed, that of
the significance of the results.  How can we tell when the communities
detected by one method or another are truly significant and when they could
be merely the consequence of a chance coincidence of edge positions in the
network?  Clear answers to this question are crucial if the results of
community analyses are to carry any real weight.

The modularity itself was originally proposed as a way of answering this
question~\cite{NG04}: a network with strong community structure will have
high modularity and hence the value of the modularity can be used as a
quality function for communities.  More recently, however, it has been
realized that this approach is insufficient.  Although it is true that
networks with strong community structure have high modularity, it turns out
that not all networks with high modularity have strong community structure.
Indeed, there exist networks that most observers would consider to have no
community structure at all that nonetheless have high modularity.
Guim\'era~\etal~\cite{GSA04} showed numerically that divisions exist of
ordinary random graphs that have high modularity, even in the limit of
large network size, a result confirmed in later analytic calculations by
Reichardt and Bornholdt~\cite{RB07}.  The reason for this at first peculiar
finding is actually quite straightforward: the number of possible divisions
of a network increases extremely fast with network size (faster than any
exponential), so that although it is highly improbable that any one
division will, purely by chance, have high modularity, it is, in the limit
of large size, very likely that such a division will exist among the
enormous number of possible candidates.  As a result, high modularity is
only a necessary but not sufficient condition for significant community
structure.

Several authors have suggested that instead we should look for divisions of
a network that have significantly higher modularity than the random
graph~\cite{GSA04,RB07}.  For example, one could optimize the modularity
for a large number of networks drawn from the random graph ensemble,
calculate the mean~$\mu$ and standard deviation~$\sigma$ of those
modularity values, and then compare the modularity~$Q$ of the optimal
division of the real network to those values, calculating, for instance, a
$z$-score:
\begin{equation}
z = {Q-\mu\over\sigma},
\label{eq:zscore}
\end{equation}
which measures how many standard deviations the real modularity is above
the mean for the random graph.  If $z\gg1$ then~$Q$ is, in a precise sense,
significantly greater than the modularity of the random graph.

This approach, however, has a number of problems.  First, it can generate
both false positives and false negatives.  Some networks that do not have
strong community structure in the traditional sense nonetheless have
modularity significantly above that of the random graph, as shown for
example in~\cite{Massen2005}.  Conversely, there are also some networks
that are widely agreed to show strong community structure but whose
modularity is \emph{not} significantly greater than the random graph.  We
give some examples of this type of behavior later in this paper.  (To be
fair, such examples appear to be rare, so that a large difference in
modularities may in some situations be considered supporting, though not
conclusive, evidence of community structure.)

More importantly, however, the difference in modularities does not really
address the question we want to answer.  In this paper we argue that the
defining property of significant community structure is not a high
modularity, but a community structure that is robust against small
perturbations of the network.  If a small change in the network---an edge
added here, another deleted there---can completely change the outcome of
our community finding calculations then, we argue, the communities found
should not be considered trustworthy.  The $z$-score is not, in general, a
good measure of this type of robustness or fragility in a network, but
there exist other measures that, as we will show, appear to work well.

\section{Robustness of community structure}
\label{sec:robapp}
An interesting approach to testing the significance of community
assignments has been proposed by Massen and Doye~\cite{Massen2007}, who
investigated the distribution of modularity values for a variety of
networks, both real and computer generated, using a simulated annealing
technique similar to that of Reichardt and Bornholdt~\cite{Reichardt2006}
combined with a parallel tempering scheme of the type commonly used to
equilibrate simulations of glassy systems~\cite{ED05}.  As a function of
the annealing temperature they investigated (among other things) the
average modularity of divisions found, with higher temperatures favoring
poor divisions (low modularity) and lower temperatures favoring better ones
(high modularity).

In low-temperature systems, where only states of high modularity are
sampled, they found two distinct behaviors.  In most real networks they
found that the states sampled correspond to roughly the same division of
the network into communities, while in random graphs the states sampled
correspond to a variety of quite different divisions.  This suggests that
real-world networks typically have a clear global modularity maximum with
no other competitive maxima, while random graphs have many competing
maxima.  In the language of physics, the distribution of maxima has a band
gap between the ground and excited states in the real networks, but no band
gap in the random graph.  (One can also think of the system's behavior by
analogy with glassy systems, which have many competing energy minima, and
non-glassy ones, which typically do not.  Indeed, ideas from the theory of
spin glasses, in particularly replica symmetry, have proved useful in the
study of modularity~\cite{RB07}, suggesting that the difference between the
community structure of random and real-world networks may be connected with
the phenomenon of replica symmetry breaking.)

One can make use of this observation to identify community structure of the
kind found in random graphs that occurs purely as a result of chance
fluctuations: if we observe multiple modularity maxima in a network,
corresponding to distinct community assignments and having roughly equal
height, we can conclude that the assignments in question are not
trustworthy.  This approach will reliably rule out random graphs
themselves---a basic task that any significance test must certainly be
capable of---but it can in principle also rule out other cases and does so
in a natural way, since any network that has many different community
assignments of roughly equal merit can reasonably be said not to show clear
community structure.

This approach provides only a way to rule \emph{out} candidate assignments.
It allows us firmly to reject some possibilities because of the structure
of the modularity maxima, but we can never guarantee that an observed
community assignment is significant solely on the basis of this test.
Having multiple competing modularity maxima is a good indicator that the
community structure given by the highest of those maxima is not
trustworthy, but it is also possible that chance fluctuations could produce
a network in which the highest maximum is substantially higher than any
other even if the network has no underlying community structure.  In this
respect, the method is similar to other significance tests in statistics.
Significance tests only ever reject hypotheses (or fail to reject them) but
can never absolutely confirm a hypothesis to be correct.

Massen and Doye proposed to implement tests of this kind by using their
simulated annealing method to find all or a representative subset of the
assignments having greatest modularity in a network and then see if they
have similar community structure.  Simulated annealing, however, is
computationally costly and is usually not the optimization method of
choice.  And the approach of Massen and Doye cannot easily be generalized
to other optimization methods, such as the spectral method.  We propose,
therefore, a different approach based on network perturbations.

Small changes to a network---the addition or removal of a few edges, for
example---will in general result in small changes to the value of the
modularity for particular partitions of the network.  In a network with
many closely competitive modularity maxima, this can change the relative
heights of the maxima with the result that the global optimum may shift
from one maximum to another.  In a network with only a single optimum on
the other hand this cannot happen, prevented in effect by the presence of
the band gap.  Thus, a simple way to determine whether the network we are
looking at has just a single optimum is to perturb the network slightly and
observe the resulting change in the optimal partition.

This idea is the basis for our proposed method.  In effect we turn the
question of the significance of a division of a network into a question
about the robustness of that division against perturbations, and the latter
question can in practice be answered more easily.  Our method also has the
substantial advantage of being entirely agnostic about the way we discover
our communities.  We are not even required to use a modularity optimization
technique---any technique that reliably finds community structure where
present will do.  We describe our method in detail in the following
sections.

\section{Quantification of network robustness}
\label{sec:meth}
Our approach has two key components: perturbation of the network and
quantification of the resulting change in the community structure.  We
describe these two components in turn.

\subsection{Network perturbation}
\label{sec:perturbation}
We wish to specify a method for perturbing an arbitrary network by an
arbitrary amount.  In order to make comparison of communities
straightforward, we restrict our perturbed networks to having the same
numbers of vertices and edges as the original unperturbed network---only
the positions of the edges will be perturbed.  Furthermore, we desire that
a network perturbed only a small amount has just a few edges moved, while a
maximally perturbed network becomes completely random and uncorrelated with
the original.

There are a number of ways in which this could be achieved but one of the
simplest is the following.  We define a random graph with $n$ vertices and
$m$ edges in standard fashion by distributing the edges between vertex
pairs such that the probability of any particular edge falling between
vertices~$i$ and~$j$ is~$e_{ij}/m$.  This implies that the expected number
of edges between~$i$ and~$j$ will be equal to~$e_{ij}$.  (Technically, the
diagonal elements of $e_{ij}$ are different: they are equal to \emph{twice}
the expected number of edges---the extra factor of two allows for the fact
that there are two ways of choosing a vertex pair if $i$ and $j$ are
distinct but only one way if $i$ and $j$ are equal.)

This definition still leaves us a good amount of freedom since we haven't
chosen the form of~$e_{ij}$.  Except for the constraint that the total
number of edges equals~$m$ so that $\half\sum_{ij} e_{ij}=m$, we are at
liberty to make any choice we wish, but the obvious candidate is the
so-called configuration model, which is also the null model normally used
in the definition of the modularity~\cite{NG04} and the random graph model
against which values of the modularity are usually compared~\cite{RB07}.
The expected number of edges between vertices in the configuration model is
\begin{equation}
e_{ij} = {k_i k_j\over2m},
\label{eq:cmnij}
\end{equation}
where $k_i$ is the degree of vertex~$i$ in the original network.

Now we interpolate stochastically between our original network and this
random graph by ``rewiring'' (i.e.,~moving) edges.  Specifically, we go
through each edge in the original network in turn and with
probability~$\alpha$ we remove it and replace it with a new edge between a
pair of vertices $(i,j)$ chosen randomly with probability $e_{ij}/m$.
Otherwise, with probability~$1-\alpha$, we leave the edge as it is.

If $\alpha=0$, no edges are moved and this process preserves our original
network.  If $\alpha=1$ all edges are moved and the process generates a
random graph drawn from the model ensemble.  And for values of $\alpha$ in
between it generates networks in which some of the edges retain their
original positions while others are moved to positions drawn from the
random ensemble.

With the choice~\eqref{eq:cmnij} for~$e_{ij}$, the expected number of edges
between vertices $i$ and~$j$ in our perturbed network is
\begin{equation}
e_{ij}' = (1-\alpha) A_{ij} + \alpha {k_i k_j\over2m}.
\end{equation} 
where $A_{ij}$ is an element of the adjacency matrix
\begin{equation}
A_{ij} = \left\lbrace\begin{array}{ll}
           1 & \qquad\mbox{if an edge connects node $i$ and $j$,}\\
           0 & \qquad\mbox{otherwise.}
         \end{array}\right.
\end{equation}
Then the expected degree of vertex~$i$ is
\begin{align}
\av{k_i} &= \sum_{j}e_{ij}'
   = (1-\alpha) \sum_j A_{ij} + \alpha {k_i\over2m} \sum_j k_j \nonumber\\
  &= (1-\alpha) k_i + \alpha {k_i\over2m} 2m = k_i,
\end{align}
where we have made use of $\sum_j A_{ij} = k_i$ and $\sum_j k_j = 2m$.

Thus our perturbation scheme generates networks that not only have the same
number of edges as the original, but in which the expected degrees of
vertices are the same as the original degrees~\footnote{Note that the
  perturbed network may have a small number of isolated nodes.  We do not
  discard these nodes, since that would make the perturbed network a
  different size from the original; instead we assign each isolated node to
  its own community.}.

\subsection{Quantifying differences in community structure}
\label{subsec:distances}
The second component of our calculation is the comparison of the optimal
division of the perturbed network to the optimal division of the original
network, to see if the community structure has changed significantly.  A
number of methods for measuring similarities or differences between
partitions of a network have been proposed in the past.  They can be
divided roughly into three groups: methods based on pair counting, methods
based on cluster matching, and information theoretic methods.  We begin by
reviewing some of these before we discuss our choice, the variation of
information.  Our discussion follows that of Meila~\cite{Meila2007}.

Let $C$ and $C'$ be two divisions of the same network into communities.  We
will refer to such divisions as \defn{community assignments}.

Measures of the similarity or difference between two community assignments
based on \defn{pair counting} focus on the number of pairs of vertices that
are in the same or different communities in both assignments.  Such
measures include the \defn{Jaccard coefficient} and the \defn{Rand index}.
We define the following four numbers:
\begin{align*}
a_{00} &= \mbox{pairs in different communities in both $C$ and $C'$,} \\
a_{11} &= \mbox{pairs in the same communities in both $C$ and $C'$,} \\
a_{01} &= \mbox{pairs in different (same) communities in $C$ ($C'$),} \\
a_{10} &= \mbox{pairs in same (different) communities in $C$ ($C'$).}
\end{align*}
Then, for example, the unadjusted Rand index~\cite{Rand1971} is defined to
be the ratio of the number of pairs clustered in the same way in both
assignments to the total number of pairs thus:
\begin{equation}
R(C,C') = \frac{a_{11}+a_{00}}{a_{10}+a_{01}+a_{00}+a_{11}}.
\end{equation}
The Rand index is also sometimes used in an adjusted form in which a
null-model expectation value is subtracted from the unadjusted index to
give a value that is axiomatically zero in the null model.  Such adjusted
indices have the disadvantage, however, of non-locality~\cite{Meila2005}:
the distance between two community assignments that differ only in one
region of the network depends on how the rest of the network is
partitioned.

An alternative approach is \defn{cluster matching,} as embodied in measures
such as the \defn{van Dongen metric} and the \defn{classification error}.
These measures attempt to determine the best match for each cluster in $C$
to one of the clusters in~$C'$.  Suppose our two community assignments $C$
and $C'$ are composed of $K$ and $K'$ communities respectively.  The
individual communities we will denote $C_1\ldots C_K$ and $C'_1\ldots
C'_{K'}$.  Then let $n_k$ and $n'_{k'}$ be the size of communities $C_k$
and $C'_{k'}$ and $n_{kk'}$ be the number of vertices common to communities
$C_k$ and $C'_{k'}$ (i.e.,~$n_{kk'} = |C_{k} \cap C'_{k'}|$).  Then the
normalized van Dongen metric is defined by~\cite{Dongen2000}
\begin{equation}
D(C, C') = 1 - \frac{1}{2n} \Biggl[ \sum_{k=1}^{K}\max_{k'}n_{kk'}
           + \sum_{k'=1}^{K'}\max_{k}n_{kk'} \Biggr] .
\end{equation}
Note that such measures ignore any subdivisions of a community that is
never chosen as a match to a community in the other assignment.  For
example, suppose
\begin{subequations}
\begin{align}
C   &= \bigl\lbrace \set{a,b,c}, \set{d,e,f,g} \bigr\rbrace, \\
C'  &= \bigl\lbrace \set{a,b,c}, \set{d,e}, \set{f,g} \bigr\rbrace, \\
C'' &= \bigl\lbrace \set{a,b,c}, \set{d}, \set{e}, \set{f,g} \bigr\rbrace.
\end{align}
\label{eq:example}
\end{subequations}
Under the van Dongen scheme $D(C, C')=D(C, C'')$, although many would claim
(and most other measures agree) that $C$ is more similar to $C'$ than
to~$C''$.

A third class of measures for comparing community assignments is based on
information theoretic ideas~\cite{Cover2006}.  In measures such as these,
we regard our community assignments as ``messages'' and consider the
Shannon information content of these messages.  The most common way to do
this is to define $x_i$ to be the label of the community that vertex~$i$
belongs to in $C$ and $y_i$ to be the community it belongs to in~$C'$.
Then the messages consist simply of the ordered sets $\set{x_i}$
and~$\set{y_i}$.  If one knows the joint distribution from which the $x$'s
and $y$'s are drawn one can then calculate various standard information
measures.  The usual assumption is that the joint distribution is equal
simply to that of the observed community assignment.  In other words, $x$
and $y$ are assumed to be values of random variables~$X$ and~$Y$ with joint
distribution $P(X=x,Y=y)=n_{xy}/n$, where $n$ is the total number of
vertices in the network.  This immediately implies also that $P(X=x)=n_x/n$
and $P(Y=y)=n'_y/n$.

In a slight abuse of terminology, we can then define the \defn{mutual
  information} between the assignments $C$ and $C'$ to be equal to the
mutual information between the corresponding random variables:
\begin{align}
I(C;C') &= I(X;Y) \nonumber\\
        &= \sum_{x=1}^{K} \sum_{y=1}^{K'} P(x,y)
           \log \frac{P(x,y)}{P(x)P(y)},
\label{eq:mutual}
\end{align}
where we use the shorthand notation $P(x)$ to denote $P(X=x)$ and similarly
for the other distributions.  (Within physics, researchers have
traditionally used the natural logarithm in expressions such
as~\eqref{eq:mutual}, while in computer science the logarithm base~2 is
more common.  The choice makes only the difference of a multiplicative
constant, however, and has no effect on any of our results.)

The mutual information measures how much information we learn about~$C'$ if
we know~$C$.  If $C$ and $C'$ are identical, then we learn everything
about~$C'$ from~$C$.  If they are entirely uncorrelated then we learn
nothing.  One way to express this is to make use of $P(x,y)=P(x|y)P(y)$ to
write
\begin{align}
I(X;Y) &= \sum_{xy} P(x,y) \log P(x|y) - \sum_x P(x) \log P(x) \nonumber\\
       &= H(X) - H(X|Y),
\end{align}
where $H(X)$ is the information (or entropy) of~$X$ and $H(X|Y)$ is the
conditional entropy, i.e.,~the additional information needed to
describe~$X$ once we know~$Y$.  Thus if $Y$ tells us nothing about~$X$ the
two terms are equal and $I(X;Y)$ is zero.  In essence the mutual
information tells us the same thing as the conditional entropy, but the
mutual information is symmetric in $X$ and~$Y$ where the conditional
entropy is not, which makes the former a more attractive measure of
distance than the latter.

The mutual information alone, however, is not a good measure of the
difference between our community assignments.  Consider, for example, the
three example assignments of Eq.~\eqref{eq:example}.  In this case the
conditional entropies $H(C|C')$ and $H(C|C'')$ are both zero, because given
the community assignments $C'$ and $C''$ (and the appropriate mapping of
community labels from one assignment to the other) we can deduce the
assignment~$C$.  (The mapping of labels must be given, since the labels are
arbitrary and we do not want our measure to register a difference between
two assignments that in fact differ only in a permutation of the labels.)
Therefore $I(C,C') = I(C,C'') = H(C)$ in this case, which is clearly not a
useful answer.  This problem is usually dealt with by normalizing the
mutual information.  There are a number of ways of accomplishing this but,
for example, one can define
\begin{equation}
  I_\mathrm{norm}(C,C') = \frac{2I(C,C')}{H(C)+H(C')}.
\end{equation}
A variant of this measure has been used by Danon~\etal~\cite{Danon2005} to
define standardized tests for the performance of community finding
algorithms.  Although the measure works, it is quite difficult to
interpret, particularly in the normalized form, which makes it hard to give
a simple statement about what the values mean (other than to say they get
larger as community assignments become more similar).

\subsection{Variation of information}
In our work we make use of a different information theoretic measure, the
\defn{variation of information}~\cite{Meila2007,Meila2005,Meila2002}.  The
variation of information is defined by
\begin{align}
V(C,C') &= V(X,Y) \nonumber\\
        &= H(X) + H(Y) - 2I(X;Y) \nonumber\\
        &= H(X|Y) + H(Y|X) \nonumber\\
        &\hspace{-3em} = - \sum_{xy} P(x,y) \log \frac{P(x,y)}{P(y)}
           - \sum_{xy} P(x,y) \log \frac{P(x,y)}{P(x)} .
\end{align}
The variation of information is the sum of the information needed to
describe $C$ given $C'$ and the information needed to describe $C'$
given~$C$.  It has a number of desirable properties that other measures
lack.  It is a true metric on the space of community assignments, having
all the properties of a proper distance measure.  It is also a local
measure in the sense described above and it returns the intuitively correct
answer for the example of Eq.~\eqref{eq:example}, that $V(C,C'')>V(C, C')$.

The maximum value of the variation of information is $\log n$, which is
achieved when the community assignments are as far apart as possible, which
in this case means that one of them places all the nodes together in a
single community while the other places each node in a community on its
own.  The maximum value increases with $n$ because larger data sets contain
more information, but if this property is undesirable one can simply
normalize by $\log n$, as we do in the calculations presented here.  In
fact, since we will always be comparing networks of the same size, the
normalization is irrelevant anyway.

\section{Methods}
\label{sec:impl}
We now have all the components we need to describe our method as applied to
a given network.  First, we find the community assignment~$C$ that
maximizes the modularity of the network, or the best approximation to it
given the optimization algorithms available.  Second, we perturb the
network as described in Section~\ref{sec:perturbation} to create a new
network, find the optimal community assignment~$C'$ for that perturbed
network, and measure the variation of information between $C'$ and~$C$.  We
repeat this second step many times to derive an average value for the
variation of information, and repeat the entire calculation for a range of
different values of the perturbation parameter~$\alpha$.

For comparison, we also perform the same set of calculations on a random
graph drawn from a configuration model with the same degree sequence as the
original network.  Then we repeat the process for several more such random
graphs and average the values of the variation of information.

The computer time required to complete the calculations depends on the
method used to optimize the modularity, the number of random graph samples
taken, and the number of different values of~$\alpha$.  In our
calculations, as mentioned above, we use the spectral optimization method
of~\cite{Newman06b}, which is reasonably fast, though certainly not the
fastest available, and average over 10 or 100 random graphs depending on
network size for each of 40 different values of~$\alpha$ from 0 to~1.  The
complete calculation for the largest network studied here, with nearly 5000
vertices, took about a day on a standard desktop computer.

\section{Results}
\label{sec:results}
As a first demonstration of the method, we have applied it to a set of
computer generated networks of a type proposed in~\cite{GN02} and used
widely in the evaluation of community detection algorithms.  These networks
consist of $128$ vertices divided into $4$ communities of $32$ nodes each.
Each vertex pair is connected by an edge with one of two different
probabilities, one for pairs in the same group and one for pairs in
different groups, with values chosen so that the expected degree of each
vertex remains fixed at 16.  As the average number~$b$ of between-group
connections per vertex is increased from zero, the community structure in
the network, stark at first, becomes gradually obscured until, at the point
where between- and within-group edges are equally likely, the network
becomes a standard Poisson random graph with no community structure at all.

\begin{figure}
\includegraphics[width=8cm]{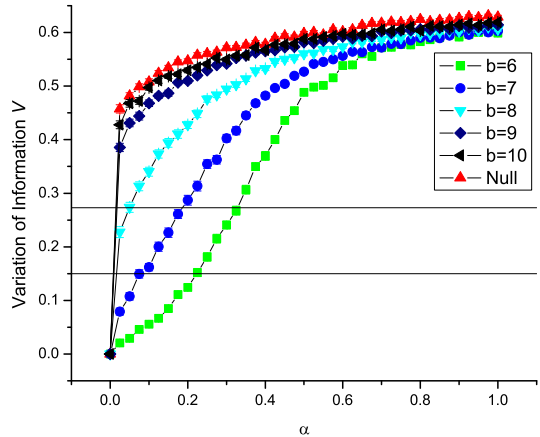}
\caption{The variation of information as a function of the perturbation
  parameter~$\alpha$ for the $128$-node four-community test networks
  described in the text ($100$ networks per point).}
\label{fig:ztest}
\end{figure}

Figure~\ref{fig:ztest} shows the results of the application of our analysis
method to graphs of this type.  The figure shows the value of the
normalized variation of information as a function of the parameter~$\alpha$
that measures the amount of perturbation.  As we can see, the variation of
information starts at zero when $\alpha=0$, as we would expect for an
unperturbed network, rises rapidly, then levels off as $\alpha$ approaches
its maximum value of~1.  Also shown is the curve for a random graph null
model of the type described above.

For large values of~$b$, such as $b=10$, the curve of the variation of
information is essentially identical to that of the null model, indicating
that whatever community structure has been found by the algorithm is no
more robust against perturbation than that of a random graph.  But as $b$
gets smaller the variation of information increases slower as a function
of~$\alpha$ and the curves depart significantly from the null model,
indicating that the community structure discovered by the algorithm is
relatively robust against perturbation.

As an aid to the interpretation of the results, we have also included in
the figure (and in all subsequent similar figures) horizontal lines
corresponding to the value the variation of information would take if we
were to randomly assign $10\%$ and $20\%$ of the vertices to different
communities.  The fact that the curves of variation of information cross
these lines at larger values of $\alpha$ in some cases than others
indicates that the community structure is more or less robust to
perturbation.  Indeed, one could simply quote the values of $\alpha$ at
which the crossings occur as a single scalar measure of robustness, but to
do so can mean missing interesting structure present in the full curves, so
we have avoided this approach in our calculations.

\begin{figure*}
\begin{center}
\hfill
\subfigure[~Zachary's karate club]{%
\includegraphics[width=7cm]{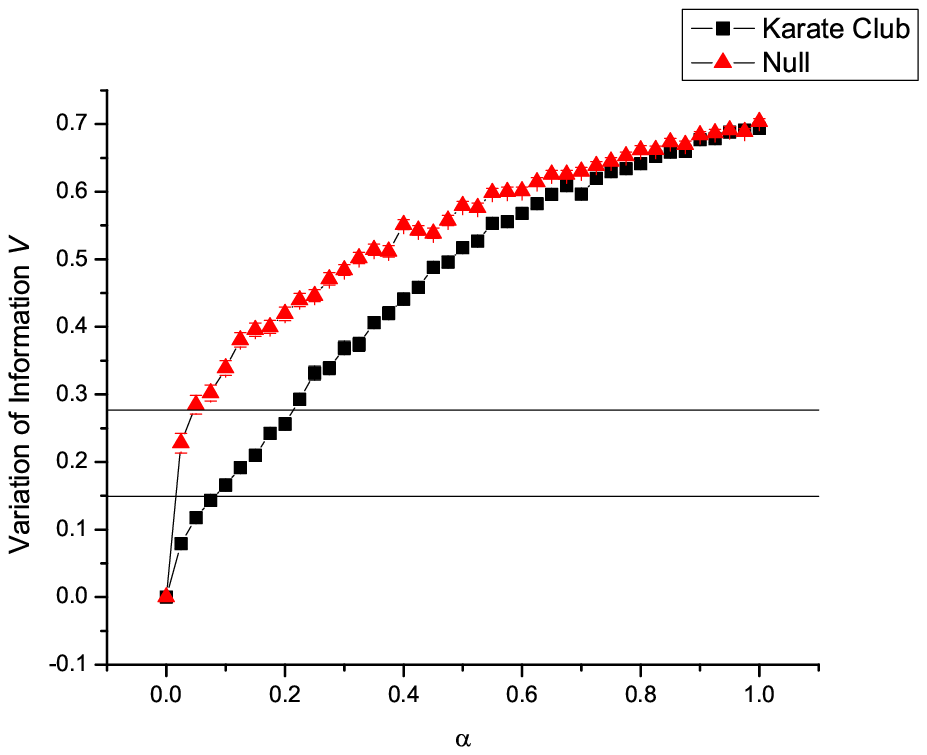}}
\hfill
\subfigure[~Social network of positive sentiments]{%
\includegraphics[width=7cm]{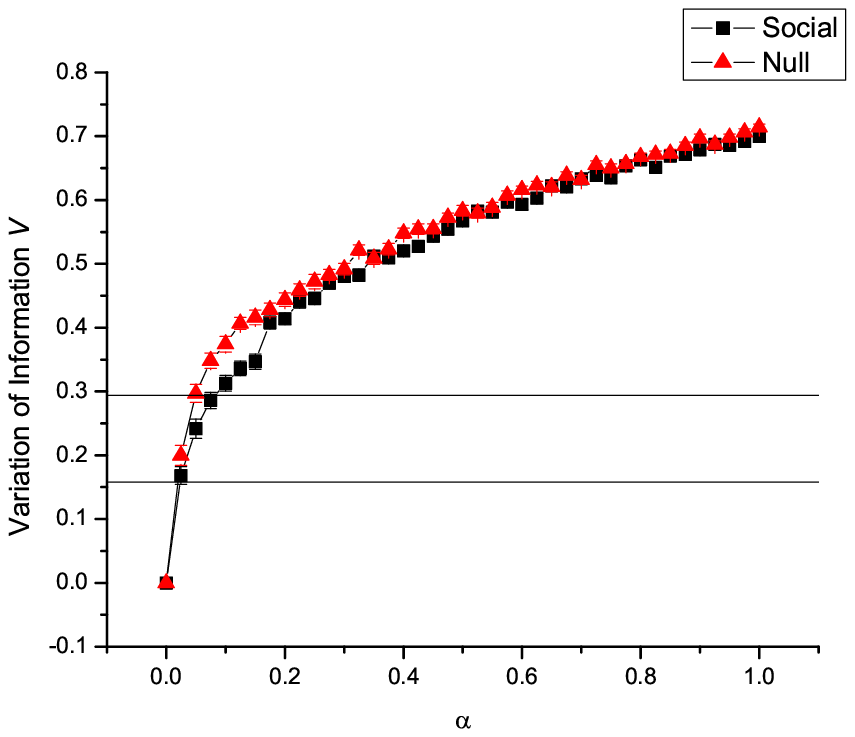}}
\hfill\null\\
\hfill
\subfigure[~Protein structure network]{%
\includegraphics[width=7cm]{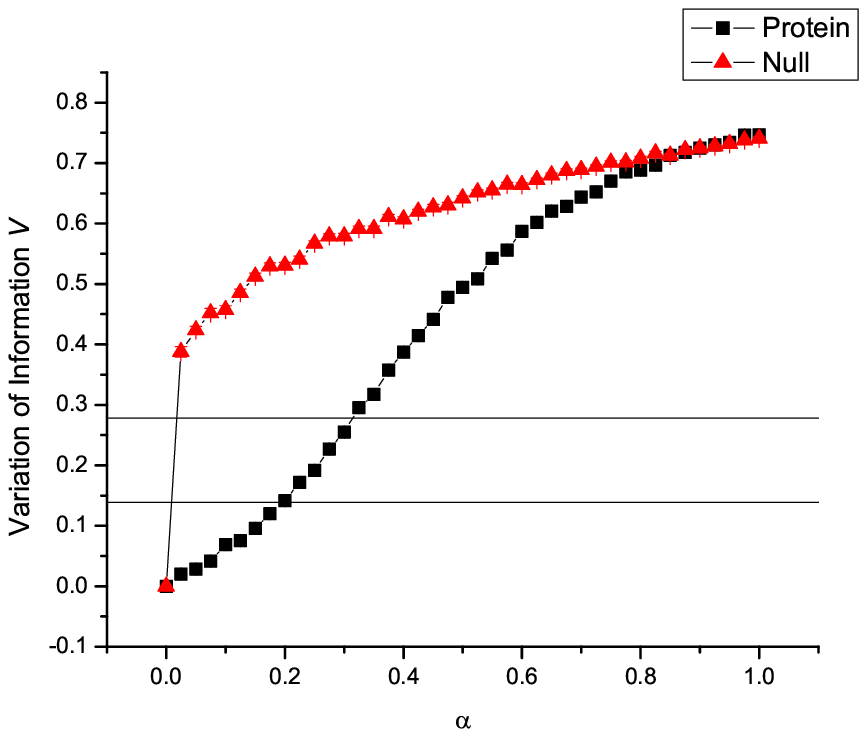}}
\hfill
\subfigure[~Metabolic network of \textit{C. Elegans}]{%
\includegraphics[width=7cm]{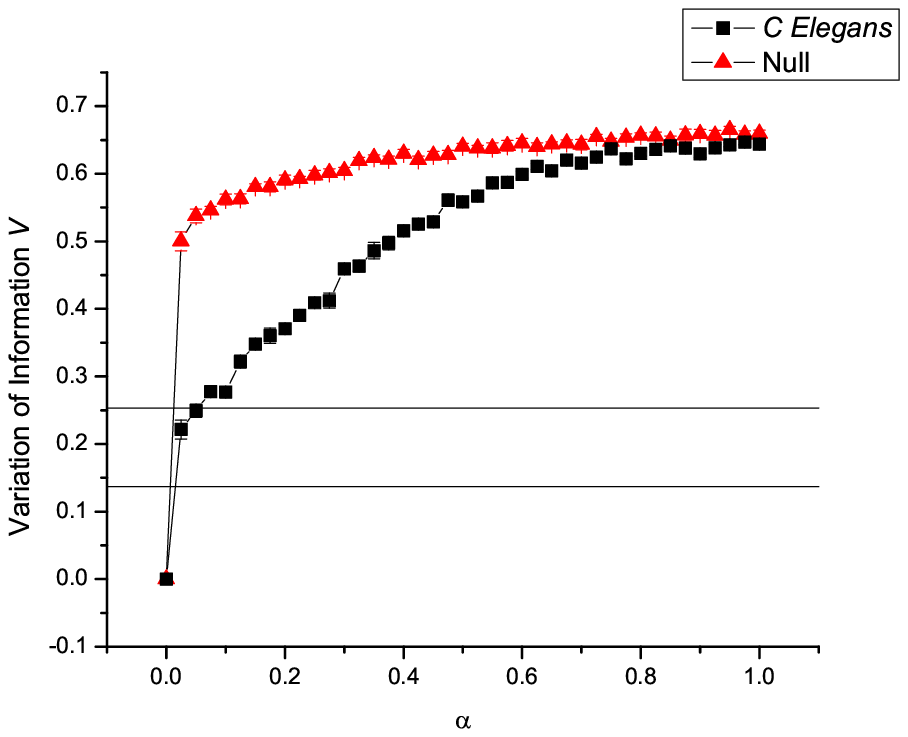}}
\hfill\null\\
\hfill
\subfigure[~Electronic circuit]{%
\includegraphics[width=7cm]{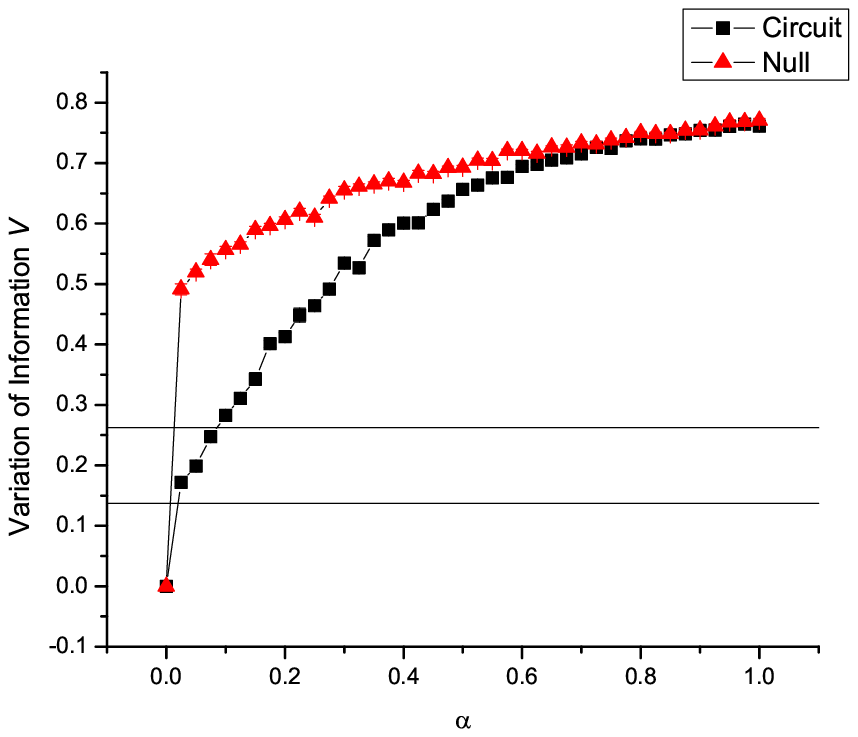}}
\hfill
\subfigure[~Power grid]{\includegraphics[width=7cm]{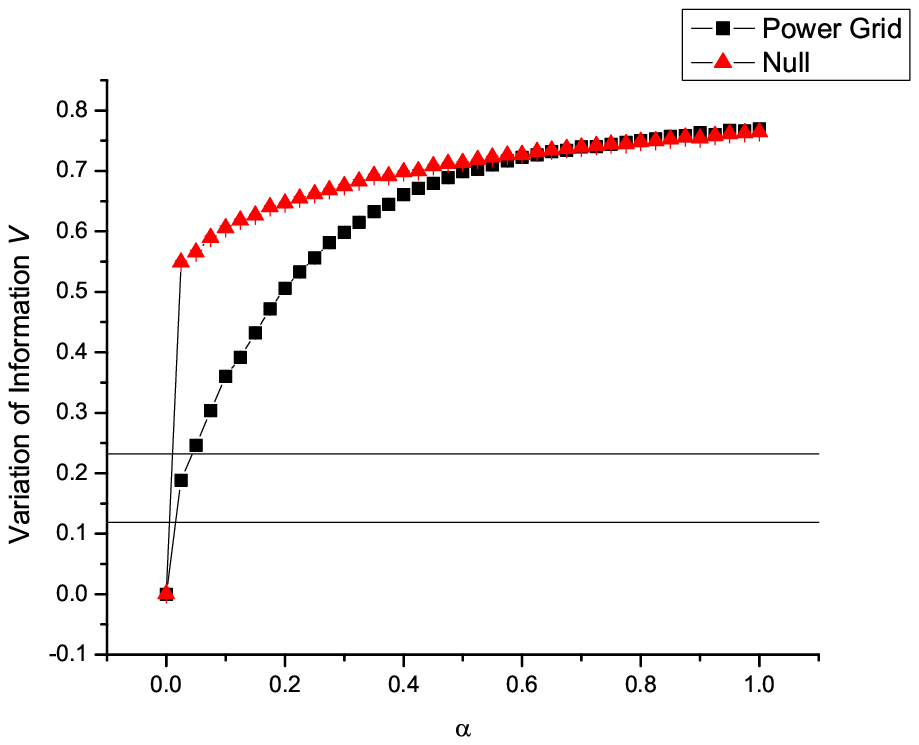}}
\hfill\null
\end{center}
\caption{The variation of information as a function of the perturbation
  parameter~$\alpha$ for six real-world networks as described in the text,
  along with equivalent results for the corresponding random graphs.}
\label{fig:results}
\end{figure*}

Turning now to real-world networks, we have tested our method on a variety
of examples including social, technological, and biological networks.  A
selection of results are shown in Fig.~\ref{fig:results}.  Some summary
statistics for the same networks are given in Table~\ref{tab:results}.

Figure~\ref{fig:results}a shows the curve of variation of information as a
function of~$\alpha$ for one of the best studied examples of community
structure in a social network, the ``karate club'' network of
Zachary~\cite{Zachary77}.  (The karate club has become so common an example
in this context that it has almost come to the point where no publication
about community structure could be complete if it failed to discuss this
network.)  The vertices in this network represent members of a karate club
at a US university in the 1970s and the edges represent friendship between
members based on independent observations by the experimenter.  The network
is widely believed to show strong community structure and repeated studies
have upheld this view.

The black points (squares) in the figure show the variation of information
for the real network while the red points (triangles) show the results for
the equivalent random graph.  It is clear in this case that the community
structure discovered in the real network is substantially more robust
against perturbation than that of the random graph.  For example, the curve
for the real network crosses the line representing reassignment of 20\% of
the vertices close to the point where $\alpha=0.2$.  Speaking loosely, we
can say that about 20\% of the edges must be rewired before 20\% of the
vertices move to different communities.  For the random graph, on the other
hand, only about 5\% of the edges need be rewired to reach this point.

A contrasting situation is seen in Fig.~\ref{fig:results}b, which shows
results for another social network, a network of friendships among a group
of first-year university students at the University of Groningen in the
Netherlands~\cite{Duijn03}.  Data for this network were collected by
circulating questionnaires among members of the group; edges between pairs
of students indicate that at least one member of the pair stated either
that they were friends or that they had a ``friendly relationship.''
Despite the similar nature of this network and the karate club network
(both are networks of friendship among university students), the results of
the analysis are quite different.  In the Groningen network, as
Fig.~\ref{fig:results}b shows, there is essentially no difference between
the variation of information for the real network and the corresponding
random graph.  The community structure algorithm does detect some structure
in the network, finding four communities of sizes $5$, $7$, $9$, and~$11$
vertices respectively and a respectable modularity score of 0.368, but our
robustness analysis indicates that this structure is not significant and
therefore should probably not be taken as indicative of the presence of any
real communities in the network.

Our next two examples are both biological networks.  The first
(Fig.~\ref{fig:results}c) represents the structure of a protein (an
immunoglobin), with the vertices representing $\alpha$-helices and
$\beta$-sheets and an edge between any two that are less than 10\AA\
apart~\cite{Milo2004}.  The second (Fig.~\ref{fig:results}d) represents
known portions of the metabolic network of the nematode \textit{C.
  Elegans}, with vertices representing metabolites and edges representing
metabolic reactions~\cite{Jeong2000}.  Again the two networks show
contrasting behaviors.  The community structure in the protein network
displays substantial robustness against perturbations, with a wide gap
between the variation of information curves for the true network and the
random graph.  A value of the variation of information equivalent to the
randomization of 20\% of the vertices is not reached until a perturbation
strength of around $\alpha=0.3$.  The metabolic network by contrast reaches
the same point around $\alpha=0.05$, not much better than the equivalent
random graph.  The curve of variation of information for the metabolic
network does however remain distinct from that of the corresponding random
graph for higher values of~$\alpha$, indicating that some portion of the
community structure found is relatively robust.

Our last two examples are technological networks, an electronic
circuit~\cite{Brglez1989,Cancho2001} and a network representation of the
power grid of the western United States~\cite{WS98}.  Both of these
networks show weak community structure similar to that of the metabolic
network, with a variation of information that increases rapidly with
$\alpha$ at first, indicating that much of the observed structure is quite
fragile to perturbation, though the curves again remain distinct; we
conclude that the networks show some community structure, even if the
effects are not strong.

\begin{table}
\begin{center}
\setlength{\tabcolsep}{6pt}
\begin{tabular}{l|c|r}
  network             & modularity & $z$-score \\
\hline
  Test $b=6$          & 0.373 & $21.0$ \\
  Test $b=7$          & 0.311 & $11.1$ \\
  Test $b=8$          & 0.248 & $2.63$ \\
  Test $b=9$          & 0.217 & $-2.04$ \\
  Test $b=10$         & 0.210 & $-2.99$ \\
\hline
  Karate club         & 0.419 & $1.77$ \\
  University students & 0.368 & $-0.19$ \\
  Protein structure   & 0.763 & $24.5$ \\
  \textit{C. Elegans} metabolic &0.434 & $25.4$ \\
  Electronic circuit  & 0.805 & $31.2$ \\
  Power grid          & 0.925 & $100.8$ \\
\end{tabular}
\end{center}
\caption{Maximum modularity and $z$-scores for each of the networks studied
  here.  The first five lines of results are averages over
  computer-generated random networks as described in the text.  The final
  six are real-world examples.}
\label{tab:results}
\end{table}

Now compare these results with those given in Table~\ref{tab:results}.  The
final column of the table gives a $z$-score for each network calculated as
described in the introduction (see Eq.~\eqref{eq:zscore}).  The comparison
with the curves for variation of information is an interesting one.  Five
of the six networks have positive $z$-scores, but not all of the scores are
large enough to make the results statistically significant.  The most
common rule of thumb is that measurements are significant if they lie more
than two standard deviations from the mean of the null model, i.e.,~if
$z>2$.  By this rule, neither of our social networks have significant
community structure, a surprising conclusion given that it is universally
accepted that the karate club network has strong community structure,
confirmed by repeated studies using many methods, and our variation of
information calculation confirms this also.  For the network of university
students, on the other hand, the $z$-score and our calculations concur,
both indicating that the community structure found is not significant, also
a troubling result, since it implies that a low $z$-score may correspond
either to strong community structure or to none at all.

The remaining four networks all have very large $z$-scores; the smallest of
them is $24.5$ and an observation twenty-four standard deviations from the
mean will be considered significant by essentially any standard.
Curiously, however, there seems to be little correlation between the
$z$-scores and the robustness of the community structure.  The highly
robust protein structure network, for instance, has the lowest $z$-score of
the four, while the power-grid---one of the networks we concluded to have
only rather weak community structure---has a spectacular $z=100.8$.
Overall, therefore, it appears that while $z$-scores for modularity values
probably do give some indication of the strength of community structure,
they are in general unreliable and should not be trusted unless backed up
by other calculations, such as those presented here.

\section{Conclusions}
\label{sec:conc}
In this paper we have examined measures of significance for network
community structure that address the question of when communities found in
a network can be considered believable, and could not reasonably have been
the result of chance fluctuations in network structure.  We have argued
that high modularity scores, the conventional measure of significance, have
less discriminatory power than measures that quantify the robustness of
community assignments to network perturbation.  We have proposed a method
for perturbing networks and a measure of the robustness under such
perturbations based on the information-theoretic distance metric known as
the variation of information.  In applications to both real and
computer-generated example networks, our method appears able to distinguish
successfully and clearly between examples that show strong community
structure and examples that do not.

In considering future directions for research, we note that all of the
calculations presented here focus on the quality of partitions of an entire
network.  It is possible that there might be significant community
structure in one part of a network and not in another, and were this the
case one would like to be able to detect it.  The methods described here
could potentially be useful for this type of investigation: one can ask
whether some communities in a network are robust under perturbation while
others are not.  The global variation of information, however, cannot
reveal this type of distinction and more detailed local measures are
needed.  We look forward to further developments in this area.

\begin{acknowledgments}
  We thank J\"org Reichardt for useful discussions.  This work was funded
  in part by the National Science Foundation under grant DMS--0405348 and
  by the James S. McDonnell Foundation.
\end{acknowledgments}

\end{document}